\documentstyle[12pt,twoside,fleqn,espcrc1]{article}

\newcommand{\AmS}{{\protect\the\textfont2
  A\kern-.1667em\lower.5ex\hbox{M}\kern-.125emS}}
\newcommand{\nc}{\newcommand}
\nc{\be}{\begin{equation}}
\nc{\ee}{\end{equation}}
\nc{\bea}{\begin{eqnarray}}
\nc{\eea}{\end{eqnarray}}
\nc{\beas}{\begin{eqnarray*}}
\nc{\eeas}{\end{eqnarray*}}
\nc{\noi}{\noindent}
\nc{\sD}{\not \! \! D}
\nc{\s}[1]{\not \! #1}
\nc{\non}{\nonumber}
\nc{\bb}{\bibitem}
\nc{\lf}{\left}
\nc{\mb}[1]{\makebox[#1]{}}
\nc{\pa}{\partial}
\nc{\sA}{\not \! \! A}
\nc{\newsec}[1]{\section{#1}\mb{0.5cm}}
\nc{\h}{\frac{1}{2}}
\nc{\ra}{\rightarrow}
\nc{\la}{\leftarrow}
\nc{\ep}{$e^+e^-\ra\pi^+\pi^-\;$}
\nc{\epp}{$e^+e^-\ra\pi^+\pi^0\pi^-\;$}
\def\mathunderaccent#1{\let\theaccent#1\mathpalette\putaccentunder}
\def\putaccentunder#1#2{\oalign{$#1#2$\crcr\hidewidth
\vbox to.2ex{\hbox{$#1\theaccent{}$}\vss}\hidewidth}}

\nc{\ti}{\mathunderaccent\tilde}
\nc{\M}{{\cal M}}
\nc{\rw}{$\rho\!-\!\omega\;$}
\title{On extracting the \rw mixing amplitude from the pion form-factor
\thanks{Talk given at the
Int. Conf. on Quark Lepton Nuclear Physics, Osaka, May 20 - 23, 1997.}
}

\author{A.G.\ Williams, 
     \address{CSSM and Department of Physics and Mathematical Physics,\\
     University of Adelaide, Australia 5005\\}
        H.B.\ O'Connell,
     \address{Department of Physics and Astronomy, University of Kentucky,\\
        Lexington, KY 40506, USA}
        and A.W.\ Thomas $^{\rm a}$       
       }

\begin{document}
\maketitle

\begin{abstract}
In this paper we improve and extend a recent analysis which showed that the
\rw mixing amplitude
cannot be unambiguously extracted from the pion electromagnetic form-factor
in a model independent way. In particular, we focus on the argument that the
extraction is sensitive to the presence of any intrinsic $\omega_I \ra\pi\pi$
coupling. Our extended analysis confirms the original conclusion, 
with only minor, 
quantitative differences. The extracted mixing 
amplitude is shown to be
sensitive to both the intrinsic coupling $\omega_I \ra\pi\pi$ and to the value
assumed for the mass of the $\rho^0$-meson.
\end{abstract}

\noindent
ADP-97-19/T256\\
UK/97-15\\
hep-ph/9707253

\section{Introduction}

The \rw mixing amplitude is traditionally extracted from the pion
electromagnetic (EM) form-factor, $F_\pi(q^2)$, as measured in \ep{}
\cite{review}.  The non-perturbative strong interaction effects that
produce the significant enhancement in the interaction around
$\sqrt{q^2}\simeq 750$MeV have been successfully parametrised using the
vector meson dominance (VMD) model.  
Our interest lies in the appearance of the isoscalar $\omega$ meson in the
isovector, $\rho^0$ meson dominated,
pion form-factor.  The traditional treatment
neglects any intrinsic coupling of the $\omega$ to the two pion final state
(i.e., that {\em not} proceeding through a $\rho^0$ meson)
and hence assumes that
the \rw mixing amplitude is purely real.  An argument due to Renard and others
\cite{renard} shows that, within certain approximations, the imaginary part of
the \rw mixing amplitude is cancelled by the intrinsic $\omega$ decay
contribution to the pion form-factor.  This argument was critically questioned
recently \cite{MOW} and argued to be unjustified. In the following we
discuss results from Ref.~\cite{OTW97}, which extend and improve
these arguments.
We confirm the central conclusion, finding, in addition, a considerable 
sensitivity to the value assumed
for the $\rho$ mass.

\section{The isospin pure basis}

Analysis of \rw mixing starts with the ``isospin pure" fields,
$\rho_I$ and $\omega_I$. These are by definition exact 
eigenstates of isospin.  The
photon to hadron interaction can be conveniently described 
using a formalism in which the renormalised vector
meson propagator is a 
$2\times2$ diagonal matrix. For the pion EM
form-factor, in the pure isospin limit, we have
no \rw mixing and no direct $\omega_I\ra\pi\pi$ coupling and hence
\be
F_\pi=\frac{1}{e}(f_{\gamma\rho_I},\,f_{\gamma\omega_I})
\left(\begin{array}{cc}
D^I_{\rho\rho} & 0 \\
0 & D^I_{\omega\omega}
\end{array}\right)
\left(\begin{array}{c}
g_{\rho_I\pi\pi}\\
0\end{array}\right),
\ee
where $e\equiv|e|$, 
$f_{\gamma V_I}$ is the coupling of the vector meson to the
photon and $g_{\rho_I\pi\pi}$ is the coupling of the $\rho$ to the two pion
final state. Here $D_{\rho\rho}^I$ and $D_{\omega\omega}^I$ are the scalar
parts of the renormalised propagators for the isospin pure fields.
As is standard in traditional treatments, 
coupling to conserved currents is assumed, allowing us to
use $D_{\mu\nu}(q^2)=-g_{\mu\nu}D(q^2)$,
where $D(q^2)$ is the
``scalar propagator." 

The two eigenstates ``mix" through the
isospin violating mixing self-energy, $\Pi_{\rho\omega}(q^2)$, 
to allow the decay
$\omega_I\ra \rho_I\ra \pi\pi$ through off-diagonal terms in the dressed,
isospin-violating vector meson propagator matrix \cite{OPTW}.  This introduces
isospin violation (IV)
in the vector meson propagator and we have (retaining only first order
in isospin violation) \cite{OPTW}
\bea\non
D^I=\left(\begin{array}{cc}
D^I_{\rho\rho} & 0 \\
0 & D^I_{\omega\omega}
\end{array}\right)\ra
D^I\!\!\!&=&\!\!\!\left(\begin{array}{cc}
D^I_{\rho\rho} & D^I_{\rho\omega}(q^2) \\
D^I_{\rho\omega}(q^2) & D^I_{\omega\omega}
\end{array}\right) \\
\!\!\!&=&\!\!\!
\left(\begin{array}{cc}
D^I_{\rho\rho} & D^I_{\rho\rho}\Pi_{\rho\omega}(q^2)D^I_{\omega\omega} \\
D^I_{\rho\rho}\Pi_{\rho\omega}(q^2)D^I_{\omega\omega} & D^I_{\omega\omega}
\end{array}\right)
+{\cal O}(({\rm IV})^2).
\eea
However, {\it a priori}, in an effective Lagrangian model involving
the vector mesons, all isospin violation sources are {\it equally likely}
and there is no reason to exclude  the possibility of
the ``intrinsic decay", $\omega_I\ra \pi\pi$, through the
coupling $g_{\omega_I\pi\pi}$.  
The appropriate VMD-based expression for the pion form-factor in the isospin
pure basis would then be
given by
\bea\non
{F_\pi(q^2)}&=&\frac{1}{e}(f_{\gamma\rho_I},\,f_{\gamma\omega_I})
\left(\begin{array}{cc}
D^I_{\rho\rho} & D^I_{\rho\rho}\Pi_{\rho\omega}D^I_{\omega\omega} \\
D^I_{\rho\rho}\Pi_{\rho\omega}D^I_{\omega\omega} & D^I_{\omega\omega}
\end{array}\right)
\left(\begin{array}{c}
g_{\rho_I\pi\pi}\\
g_{\omega_I\pi\pi}\end{array}\right) \\ \non
&=&\frac{f_{\gamma\rho_I}}{e}\frac{1}{q^2-m^2_\rho(q^2)}g_{\rho_I\pi\pi}+
\frac{f_{\gamma\omega_I}}{e}\frac{1}{q^2-m^2_\omega(q^2)}\Pi_{\rho\omega}(q^2)
\frac{1}{q^2-m^2_\rho(q^2)}g_{\rho_I\pi\pi} \\
&&+\frac{f_{\gamma\omega_I}}{e}
\frac{1}{q^2-m^2_\omega(q^2)}g_{\omega_I\pi\pi},
\label{one}
\eea
where a fourth term on the RHS involving both $\Pi_{\rho\omega}$ and
$g_{\omega_I\pi\pi}$ has been neglected since it is second order in isospin
violation.
One is always free to consider models where $g_{\omega_I\pi\pi}$
is strictly zero, but there is no model-independent requirement that this 
be so.
For the renormalised, 
isospin-pure propagators, $D^I_{VV}$, we have used the 
physical $\rho$ and $\omega$ propagators, since $D_{VV}=D^I_{VV}+
{\cal O}(({\rm IV})^2)$ \cite{OPTW} and since again we are working only to
first order in isospin violation. For the physical $\rho$ and $\omega$
propagators we will use here the usual Breit Wigner form with
a momentum dependent width, i.e., $D_{VV}=\frac{1}{q^2-m^2_V(q^2)}$,
where $m_V^2(q^2)=\hat{m}_V^2-i\hat{m}_V\Gamma_Vh_V(q^2)$
and where we have defined the momentum-dependent width $\Gamma_V(q^2)
=\Gamma_Vh_V(q^2)$ with
$h_V(\hat{m}_V^2)\equiv 1$ and where the mass parameter ($\hat{m}_V$)
and width parameter ($\Gamma_V$) are
real.

\section{The physical basis}
The pion form-factor
has two distinct resonance poles, yet Eq.~(\ref{one}) has a rather complicated
pole structure. Therefore,
one might change to a basis
that has only a sum of simple poles. This can be done by transforming
to what is referred to as the ``physical" basis.
It has been traditional to assume that $\Pi_{\rho\omega}$ is a constant
which allows the off-diagonal terms in the propagator
to be removed by a rotation of the fields. 
However, recent work has argued that, when \rw mixing is generated only 
through
vector mesons coupling to conserved currents, this amplitude is necessarily
momentum dependent \cite{OPTW}. 
This result is consistent with an earlier study 
which concluded that momentum independent mixing was
acceptable for scalar particles, but not for vectors 
coupling to conserved currents \cite{CS}. 
Various demonstrative model calculations support this \cite{calcs}. 
To make allowance for this,
Maltman, O'Connell and Williams (MOW) 
defined the physical basis through {\em two} isospin
violating parameters \cite{MOW},
$\rho=\rho_I-\epsilon_1 \omega_I$,
$\omega=\omega_I+\epsilon_2 \rho_I$.
This defines a transformation matrix, $C$, via
\be
\left(\begin{array}{c}
\rho \\
\omega\end{array}\right)
\equiv C
\left(\begin{array}{c}
\rho_I \\
\omega_I\end{array}\right)
=\left(\begin{array}{cc}
1 & -\epsilon_1 \\
\epsilon_2 & 1\end{array}\right)
\left(\begin{array}{c}
\rho_I\\ \omega_I\end{array}\right).
\ee
The off-diagonal element of the physical propagator
was then defined in terms of
the physical fields in the usual way through
\be
D^{\mu\nu}_{\rho\omega}(q^2)
=i\int d^4x e^{iq\cdot x}\langle0|T(\rho^\mu(x)\omega^\nu(0))|0\rangle.
\label{kim}
\ee
As the vector mesons couple to conserved currents, we can replace
$D^{\mu\nu}(q^2)$ by $-g^{\mu\nu}D(q^2)$. Eq.~(\ref{kim}) leads us to
\bea\non
D_{\rho\omega}(q^2)&=&D_{\rho\omega}^I(q^2)-\epsilon_1D^I_{\omega\omega}(q^2)
+\epsilon_2D^I_{\rho\rho}(q^2) \\
&=&D_{\rho\rho}^I(q^2)[\Pi_{\rho\omega}(q^2)-
\epsilon_1(D_{\rho\rho}^I(q^2))^{-1}+\epsilon_2
(D_{\omega\omega}^I(q^2))^{-1}]D_{\omega\omega}^I(q^2).
\eea
Note that $D=C^{}D^IC^{T}\;.$
The physical basis is defined by requiring that there are no resonance
poles in  $D_{\rho\omega}(q^2)$, i.e., 
we choose $\epsilon_1$ and $\epsilon_2$ such that 
$D_{\rho\omega}(q^2)$ has no pole. The possible pole positions are
the vector meson pole positions, $m_\rho^2$ and $m_\omega^2$. These
singularities can be removed
if the numerator of $D_{\rho\omega}(q^2)$ vanishes at these 
positions, leading us to,
\be
\epsilon_1=\frac{\Pi_{\rho\omega}(m_\omega^2)}{m_\omega^2-m_\rho^2},\,\,\,
\epsilon_2=\frac{\Pi_{\rho\omega}(m_\rho^2)}{m_\omega^2-m_\rho^2}.
\ee
It should be noted that $m^2_\rho$ and $m^2_\omega$
are the  complex resonance pole
positions. In the
case where the mixing is momentum dependent we have
$\epsilon_1\neq\epsilon_2$ 
such that the matrix $C$ is not orthogonal and the
transformation between bases is then not a simple rotation.
In a similar fashion
to Eq.~(\ref{kim}), the coupling constants in the physical basis
are defined via
\be
\begin{array}{cc}
f_{\gamma\omega}
=f_{\gamma\omega_I}+\epsilon_2f_{\gamma\rho_I},&
g_{\omega\pi\pi}
=g_{\omega_I\pi\pi}+\epsilon_2g_{\rho_I\pi\pi} \\
f_{\gamma\rho} 
= f_{\gamma\rho_I}-\epsilon_1f_{\gamma\omega_I},&
g_{\rho\pi\pi} 
=g_{\rho_I\pi\pi}.
\end{array}\ee
Note that $\epsilon_1g_{\omega_I\pi\pi}$ is {\em second order} in
isospin violation and so is not retained.

In this way, MOW defined the form-factor in the physical basis by
\be
F_\pi(q^2)=\frac{1}{e}\left[g_{\omega\pi\pi}D_{\omega\omega}f_{\gamma\omega}
+g_{\rho\pi\pi}D_{\rho\rho}f_{\gamma\rho}+
g_{\rho\pi\pi}D_{\rho\omega}f_{\gamma\omega}\right]
+{\rm background}.
\label{kimff}
\ee
The non-resonant term, $D_{\rho\omega}$, was then included in the
background, along with 
any other non-pole terms of the Laurent series expansions
of the propagators, to arrive at an expression for the form-factor,
\be
F_\pi(q^2)=H(\epsilon_1,\epsilon_2)
\left[P_\rho+A(\epsilon_1,\epsilon_2)
e^{i\phi(\epsilon_1,\epsilon_2)}P_\omega\right]+{\rm background}.
\label{mowff}
\ee
Here $P_{\rho,\omega}$ are simple poles
and $H$ is an overall constant. The quantities $H$, $A$ and the Orsay phase,
$\phi$, can be read off from Eq.~(\ref{kimff}) as we will 
show later.
In terms of the isospin-pure basis  and the
transformation matrix, $C$, Eq.~(\ref{kimff}) can be written,
\be
F_\pi=\frac{1}{e}(f_{\gamma\rho_I}\,\,\,f_{\gamma\omega_I})C^{T}CD^IC^TC
\left(\begin{array}{c}
g_{\rho_I\pi\pi}\\
g_{\omega_I\pi\pi}\end{array}\right).\label{fpieq}
\ee
Because of the
closeness of the $\rho$ and $\omega$ pole positions,
there is no practical distinction in their analysis between the
two terms [i.e., $\Pi_{\rho\omega}(m_\rho^2)$
and $\Pi_{\rho\omega}(m_\omega^2)$], so we define a single parameter,
$\epsilon \equiv
\epsilon_2=\Pi_{\rho\omega}(m_\rho^2)/(m_\omega^2-m_\rho^2)\simeq\epsilon_1$,
where $\Pi_{\rho\omega}(m_\rho^2)\simeq\Pi_{\rho\omega}(m_\omega^2)
\simeq\Pi_{\rho\omega}(\hat{m}_\rho^2)\simeq
\Pi_{\rho\omega}(\hat{m}_\omega^2).$
In other words, from this point on we will assume (as is done in all
standard treatments) that the momentum dependence
of $\Pi_{\rho\omega}(q^2)$ is negligible {\em in the vector meson 
resonance region}. 

\section{The Renard Argument}
The major conclusion of the MOW analysis \cite{MOW}
concerns the competition between
the two sources of isospin violation, namely, $\Pi_{\rho\omega}$ and 
$g_{\omega_I\pi\pi}$. All isospin violation in
the pion form-factor is usually attributed to \rw mixing \cite{CB}, because
the intrinsic decay of the $\omega$
is assumed to be cancelled \cite{renard}. 
Consider the $\omega$ pole term of Eq.~(\ref{mowff}). The coupling
of the physical $\omega$ to the two pion final state is given by,
$g_{\omega\pi\pi}=g_{\omega_I\pi\pi} +\epsilon g_{\rho_I\pi\pi}$.
It is certainly a reasonable approximation to
assume that
the two pion intermediate
state saturates the imaginary part of $\Pi_{\rho\omega}(q^2)$ around
the $\rho$ resonance region
and that this is proportional to the two pion piece of the
$\rho$ self energy (which dominates the imaginary piece
of the total $\rho$ self energy, due to the strong $\rho\ra\pi\pi$ decay).
We have then
\be {\rm Im}\;\Pi_{\rho\omega}(m_\rho^2)=
{\rm Im}\;\Pi_{\rho\omega}^{\pi\pi}(m_\rho^2)
=\frac{g_{\omega_I\pi\pi}}{g_{\rho_I\pi\pi}}
{\rm Im}\;\Pi_{\rho\rho}^{\pi\pi}(m_\rho^2)\equiv
-G\hat{m}_\rho\Gamma_\rho h_\rho(m_\rho^2)
\simeq -G\hat{m}_\rho\Gamma_\rho\;,
\ee
where we have defined
$G\equiv\frac{g_{\omega_I\pi\pi}}{g_{\rho_I\pi\pi}}$,
and assumed $h_\rho(m^2_\rho)\simeq h_\rho(\hat{m}^2_\rho)\equiv 1$.
We hence define
\be
\Pi_{\rho\omega}(m_\rho^2)\equiv \widetilde{\Pi}_{\rho\omega}(m_\rho^2) -i
G\hat{m}_\rho\Gamma_\rho. \label{tildepi}
\ee
Assuming saturation of the absorptive part by the two pion state,
as described above, implies that
$\widetilde{\Pi}_{\rho\omega}(m_\rho^2)$ is real. 
Note that while the three pion state is kinematically
accessible it also requires isospin violation and is, in addition,
suppressed by the smaller phase space available. We follow
standard practice and ignore it here. 

Substituting Eq.~(\ref{tildepi})
into the expression for $\epsilon$, we have
\be
\epsilon = \frac{\widetilde{\Pi}_{\rho\omega}
(m_\rho^2)}{m_{\omega}^2-m_{\rho}^2}
-iG\frac{\hat{m}_\rho\Gamma_\rho}{m_{\omega}^2-m_{\rho}^2}
\equiv -iz\widetilde{T}-zG,
\;\;{\rm where}\;\;
z\equiv\frac{i\hat{m}_\rho\Gamma_\rho}{m_\omega^2-m_\rho^2},\,\,\,\widetilde{T}
\equiv\frac{\widetilde{\Pi}_{\rho\omega}(m_\rho^2)}{\hat{m}_\rho\Gamma_\rho}.
\label{black}\label{white}
\ee
Note that since in the resonance region we are neglecting momentum
dependence of the resonance widths, i.e.\ we take
$h_\omega(q^2)=h_\rho(q^2)=1$, then we see that
$m_\omega^2$, $m_\rho^2$, and $z$ are all constants. 
Recall that here, since we are using just the single parameter $\epsilon$,
the mixing amplitude $\widetilde{\Pi}_{\rho\omega}(m_\rho^2)$ should
be understood to mean $\widetilde{\Pi}_{\rho\omega}(q^2)$ with
$q^2$ in the vector resonance region.
Using this expression for $\epsilon$ in the $\omega$ pole term
gives rise to Renard's cancellation
\be
g_{\omega_I\pi\pi} +\epsilon g_{\rho_I\pi\pi}=
g_{\omega_I\pi\pi}(1-z)+ \frac{\widetilde{\Pi}_{\rho\omega}
(m_\rho^2)}{m_{\omega}^2
-m_{\rho}^2} g_{\rho_I\pi\pi}
=\left[G(1-z)+ \frac{\widetilde{\Pi}_{\rho\omega}(m_\rho^2)}{m_{\omega}^2
-m_{\rho}^2}\right] g_{\rho_I\pi\pi}
\ee
when one makes the approximation $z=1$. However, MOW find 
that $z$ has a sizeable imaginary piece \cite{MOW}
$z=0.9324+0.3511\: i$,
using the mass and width values from Bernicha {\it et al.} \cite{BCP}.
For comparison, one finds
$z=1.023+0.2038i$ using the values of Benayoun {\it et al.} \cite{Ben}.
The central observation of MOW was to point out that the
deviation of $z$ from unity leads to a substantial
contribution to $F_\pi(q^2)$ from $g_{\omega_I\pi\pi}$.

At this point
MOW neglected the $\epsilon$ dependence of the constant,
$H(\epsilon_1,\epsilon_2)$ in Eq.~(\ref{mowff}) and extracted $G$ and
$\widetilde{\Pi}_{\rho\omega}(m_\rho^2)$ by fitting $A$ and the Orsay phase
$\phi$. In order to do a more careful
analysis of the $G$ dependence of the
extracted real part of the \rw mixing amplitude,
$\widetilde{\Pi}_{\rho\omega}(m_\rho^2)$ we shall retain $H(\epsilon) \equiv
H(\epsilon_1,\epsilon_2)$, which itself has a $G$
dependence
through $\epsilon$ [via Eq.~(\ref{black})].  
Let us write Eqs.~(\ref{kimff}) and (\ref{mowff}),
[or equivalently Eq.~(\ref{one})] as
\be
F_\pi(q^2)=\frac{1}{e}[f_{\gamma\rho_I} g_{\rho_I\pi\pi} P_\rho
-f_{\gamma\omega_I}\epsilon(P_\rho-P_\omega)g_{\rho_I\pi\pi}
+f_{\gamma\omega_I} 
P_\omega g_{\omega_I\pi\pi}]\;,\label{red}
\ee
where we have used the fact that
$\epsilon(P_\rho-P_\omega)=-P_\rho\Pi_{\rho\omega}P_\omega$.
Defining the ratio
$r_I \equiv \frac{f_{\gamma\omega_I}}{f_{\gamma\rho_I}}$,
we can use Eqs.~(\ref{black}) and (\ref{white}) to write Eq.~(\ref{red}) as
\bea\non
F_\pi(q^2) &=&\frac{1}{e} f_{\gamma\rho_I}
g_{\rho_I\pi\pi}\left\{P_\rho+r_I\left[-\epsilon(P_\rho-P_\omega)
+GP_\omega\right]\right\} \\
&=&\frac{1}{e}f_{\gamma\rho_I} 
g_{\rho_I\pi\pi}\left\{P_\rho(1+izr_I\widetilde{T})+r_I\left[
-iz\widetilde{T}P_\omega
+zGP_\rho
+G(1-z)P_\omega\right]\right\}\label{g-one}
\eea
Comparing Eq.~(\ref{g-one}) to our earlier Eq.~(\ref{mowff}) it is seen to be
a straightforward matter to identify the qualities $H$, $A$ and the
Orsay phase, $\phi$. This is the central result.

Recall that to obtain this result, we followed standard treatments
and neglected the momentum-dependence of the Breit-Wigner 
widths in the resonance region,
$h_\rho(m_\rho^2)\simeq h_\rho(\hat{m}_\rho^2)=1$.
We also neglected the three pion loop contribution to \rw mixing, so that
${\rm Im}\:\Pi_{\rho\omega}(q^2)=G{\rm Im}\:\Pi_{\rho\rho}(q^2)$
and assumed a constant value for $\widetilde{\Pi}_{\rho\omega}$
in the vector meson resonance region.

We arrive at a similar conclusion 
in a more straightforward way
if we choose to stay in the isospin
pure basis throughout and simply re-arrange Eq.~(\ref{one}) to give 
\bea\non
F_\pi(q^2)=&&\frac{1}{e}\left[
f_{\gamma{\rho_I}}\frac{1}{q^2-m_\rho^2(q^2)}
+\frac{f_{\gamma{\omega_I}}}{q^2-m_\rho^2(q^2)}
\widetilde{\Pi}_{\rho\omega}(q^2)
\frac{g_{{\rho_I}\pi\pi}}{q^2-m_\omega^2(q^2)}\right. \\ 
&+&\left.\frac{f_{\gamma{\omega_I}}}{q^2-m_\omega^2(q^2)}g_{{\omega_I}\pi\pi}
\left(1-\frac{i\hat{m}_\rho\Gamma_\rho h(q^2)}{q^2-\hat{m}_\rho^2+
i\hat{m}_\rho\Gamma_\rho h(q^2)}\right)\right]
\label{pure}\eea
We see immediately that in the isospin pure treatment 
the $G$ dependence (i.e., $g_{\omega_I\pi\pi}$ dependence) is 
cancelled at $q^2=\hat{m}^2_\rho$ and 
somewhat suppressed around the pole region
(where \rw interference is most noticeable).
Including the nonresonant off-diagonal propagator in the physical basis 
would make Eqs.~(\ref{g-one}) and (\ref{pure}) identical, but as we can
only fit resonant terms
plus an unknown non-resonant background
to data, there is
no {\em practical} difference between these two equations.


\section{Results and conclusions}

To determine values for $\widetilde{\Pi}_{\rho\omega}(m_\rho^2)$ and $G$
we must choose an appropriate form factor and decide upon
Eq.~(\ref{pure}), which we re-write as
\be
F_\pi=\frac{-am_\rho^2}{q^2-\hat{m}_\rho^2+i\hat{m}_\rho\Gamma_\rho}
\left[1+r_I\frac{
\widetilde{\Pi}_{\rho\omega}(m_\rho^2)+G(q^2-\hat{m}_\rho^2)}{
q^2-\hat{m}_\omega^2+i\hat{m}_\omega\Gamma_\omega}\right].
\label{fit_form}
\ee
We now perform a fit to pion form-factor data 
to extract values for $\hat{m}_\rho$, $\Gamma_\rho$, 
$\widetilde{\Pi}_{\rho\omega}(m_\rho^2)$, $G$ and the normalisation
constant, $a$, using the fitting routine
Minuit \cite{paw} (case A in Table \ref{restit}).
We fix $m_\omega = 781.94$ MeV and $\Gamma_\omega = 8.43$ MeV, as given by
the Particle Data Group \cite{PDG}. 

It should be noted that we are here adopting this form to fit rather than
using Eqs.~(\ref{g-one}) and (\ref{mowff}), since we wish to isolate
all of the $G$-dependence
from the overall normalisation constant for the fit.  It can be seen
from Eq.~(\ref{g-one}) that $H$ in Eq.~(\ref{mowff}) is $G$-dependent
and so we can not use the fitting procedure adopted by MOW.
By construction, the constant $a$ in Eq.~(\ref{fit_form}) is independent
of $G$.
For the analysis here we will assume the SU(3) value for
$r_I\equiv g_{\gamma\omega_I}/g_{\gamma\rho_I}$, i.e., we will assume
$r_I=1/3$ \cite{DM}.  
However, from Eq.~(\ref{fit_form}) we see that all that can
really be extracted from the analysis is $r_I\widetilde{\Pi}_{\rho\omega}$
and $r_IG$ and so if another value of $r_I$ is preferred our results
can immediately be scaled in a straightforward way.
Our data set consists of the 70 points listed in Ref.~\cite{data} in the
region between 500 and 975 MeV. While the preferred value for $G$ is
quite large -- one usually expects isospin breaking at the few percent
level, rather than 10\% -- the uncertainty is such that it lies only $2
\frac{1}{2}$ standard deviations from zero.

We can investigate the importance of direct isospin violation at the 
$\omega\to\pi\pi$ vertex by imposing the condition $G=0$ (case B).
However, when fitting the $\rho$ data, it is important to remember that the
$\rho$ parameters should not be process-dependent. As noted by Benayoun 
{\it et al.} since the $\rho$ is a relatively
broad resonance the value extracted for, say, the
mass, can be greatly affected by the addition of other terms to a
phenomenological form-factor \cite{Ben}. To demonstrate that our
conclusions concerning $G$ and $\widetilde{\Pi}_{\rho\omega}$ 
do not depend on a particular
choice of mass and width from one source, we have redone the fit using the
Particle Data Group (PDG) values, which are averaged 
from a variety of
processes \cite{PDG}. 

In order to compare our 
results with the analysis of Maltman {\it et al.} \cite{MOW}, we 
perform a fit
to the same data using Eq.~(\ref{mowff})
with $H(\epsilon)$ treated as a constant (thereby absorbing it into the
normalisation constant $a$). To first order in isospin violation one has
\be
Ae^{i\phi}=\frac{r(G(1-z)-iz\widetilde{T})}{1+izr\widetilde{T}+zrG}.
\ee
The resulting fit parameters are shown in Table 1 under the heading MOW.
We see that they are very close to the fit using the full expression
Eq.~(\ref{g-one}), shown in column A. 

\begin{table}[htb]
\begin{center}
\begin{tabular}{|c|c|c|c|c|}
\hline
Parameter&A&B&PDG&MOW\\
\hline
$\widetilde{\Pi}_{\rho\omega}
(m_\rho^2)$ (MeV$^2$) & $-6832\pm 1252$&$-3844\pm271$&
$-6298\pm788$&$-6827\pm1067$\\
$G$& $0.102\pm 0.04$&$0$ (input) &$0.065\pm0.033$&$0.102\pm0.04$\\
$a$&$1.15\pm 0.02$&$1.19\pm0.01$&$1.203\pm0.004$&$1.20\pm0.01$\\
$\hat{m}_\rho$ (MeV)&$764.1\pm 0.7$&$763.7\pm0.7$&769.1 (input)
&$764.1\pm0.7$\\
$\Gamma_\rho$ (MeV)&$145.0\pm 1.7$&$146.9\pm1.4$&151.0 (input)
&$145.0\pm1.6$\\
$\chi^2/$dof&88/65&94/66&172/67&88/65\\
\hline
\end{tabular}
{\caption {Results from fitting our form-factor to data.}
\label{restit}}
\end{center}
\end{table}

In conclusion, we have seen that, in agreement with the conclusions of
Maltman {\it et al.}, the pion form factor data supports equally 
well a large range of
possible pairs of values for $G$ and $\widetilde{\Pi}_{\rho\omega}$.  In other
words, it is not possible to extract the \rw mixing amplitude in a
model-independent way.  The traditional method of extraction  
corresponds to assuming that there is no intrinsic $\omega\to\pi\pi$
coupling (i.e., that $G=0$), which is highly unlikely.  
It should also be noted that these conclusions
are entirely independent of what (if any) momentum-dependence is present
in the \rw mixing amplitude, since this was neglected in the resonance
region in the usual way.

\end{document}